\documentclass[twoside,reqno]{HERON}
\usepackage{epsfig,cite,colordvi}
\usepackage{graphicx}
\usepackage{url}
\usepackage{mathtools}
\usepackage{amssymb,amsmath,amscd,epsf}
\usepackage{times}
\usepackage{makeidx}
\usepackage{bm}
\usepackage{xcolor}
\usepackage{tipa}
\newcommand {\nc} {\newcommand}

\nc {\IR} [1]{\textcolor{red}{#1}}
\nc {\IB} [1]{\textcolor{blue}{#1}}
\nc {\IP} [1]{\textcolor{magenta}{#1}}
\nc {\IM} [1]{\textcolor{Bittersweet}{#1}}
\nc {\IE} [1]{\textcolor{Plum}{#1}}

\nc{\ninej}[9]{\left\{\begin{array}{ccc} #1 & #2 & #3 \\ #4 & #5 & #6 \\ #7 & #8 & #9 \\ \end{array}\right\}}
\nc{\sixj}[6]{\left\{\begin{array}{ccc} #1 & #2 & #3 \\ #4 & #5 & #6 \\ \end{array}\right\}}
\nc{\threej}[6]{ \left( \begin{array}{ccc} #1 & #2 & #3 \\ #4 & #5 & #6 \\ \end{array} \right) }
\nc{\half}{\frac{1}{2}}
\nc{\numberthis}{\addtocounter{equation}{1}\tag{\theequation}}
\nc{\lla}{\left\langle}
\nc{\rra}{\right\rangle}
\nc{\lrme}{\left|\left|}
\nc{\rrme}{\right|\right|}

\begin{document}
\title{Spin dependent \textit{ab initio} nonlocal No-Core Shell-Model One-Body 
Densities}

\runningheads{Preparation of Papers for Heron Press Science Series
Books}{G. Popa, M. Burrows, Ch. Elster, K.D. Launey, P. Maris, S.P. Weppner}

\begin{start}

\author{G. Popa}{1}, \coauthor{M. Burrows}{2}, \coauthor{Ch. Elster}{2}, 
\coauthor{K. D. Launey}{3}, 
\coauthor{P. Maris}{4}, \coauthor{S.P. Weppner}{5}

\index{Popa, G.}
\index{Burrows, M.}
\index{Elster, Ch.}
\index{Launey, K.D.}

\address{Ohio University, Zanesville OH 43701, USA}{1}
\address{Ohio University, Athens, OH 45701, USA}{2}
\address{Louisiana State University, Baton Rouge, LA 70803, USA}{3}
\address{Iowa State University, Ames, Iowa 50011, USA}{4}
\address{Eckerd College, St. Petersburg, Florida 33711, USA}{5}

\begin{Abstract}
Constructing microscopic effective interactions (`optical potentials') for nucleon-nucleus (NA)
elastic scattering requires in first order off-shell nucleon-nucleon (NN) scattering amplitudes between
the projectile and the struck target nucleon and nonlocal one-body density matrices. While the NN
amplitudes and the {\it ab intio} no-core shell-model (NCSM) calculations  always contain the full spin 
structure of the NN problem, one-body density matrices used in traditional microscopic folding
potential neglect spin contributions inherent in the one-body density matrix. Here we derive and
show the expectation values of the spin-orbit contribution of the struck nucleon with respect to the 
rest of the nucleus for $^{4}$He, $^{6}$He, $^{12}$C, and $^{16}$O and compare them with the scalar
one-body density matrix.
\end{Abstract}
\end{start}

\section{Introduction}
The {\it ab-initio} NCSM has considerably advanced our understanding and capability of achieving first-principles descriptions of low-lying states in light nuclear systems~\cite{Navratil:2000ww,Navratil:2000gs,Roth:2007sv,BarrettNV13,Stumpf:2015lma}, and  has over the last decade taken center stage in the development of microscopic tools for studying the structure of atomic nuclei.  
Applying this approach to nuclear reactions requires isolating important degrees of freedom, 
thus reducing the many-body to a few-body problem and
solving the latter exactly. Isolating important degrees of freedom means projecting onto a
reduced Hilbert space and thus creating effective interactions between the degrees of freedom
that are treated exactly. Since the 1960's (or earlier), such effective interactions have been
constructed by fitting relevant experimental data with usually complex functions, leading to
the well known phenomenological optical potentials (see
e.g.~\cite{Weppner:2000fi,Koning:2003zz,Furumoto:2019anr}), which are local and
energy-dependent. While the large body of phenomenological work may keep some place in
practical applications,
an overarching goal is to construct such effective interactions (optical potentials)
from the same first principles that govern
advances in many-body approaches to nuclear structure.

Starting from a multiple scattering expansion for  NA scattering, the first order term 
requires a folding integral over a nonlocal one-body density and  off-shell NN
scattering amplitudes, where the Wolfenstein amplitude A determines the central part of the folding
potential and C the spin-orbit part (see
e.g.~\cite{Chinn:1993zz,Elster:1989en,Crespo:1990zzb,Arellano:1990xu}). 
Calculations in ~\cite{Burrows:2018ggt,Gennari:2017yez} are
carried out in this spirit.  Here the non-local one-body density matrix (OBDM) and the NN scattering
amplitudes are based on the same NN interaction, leading to a consistent {\it ab initio} first order
folding effective potential. 

However, `traditional' first order folding potentials developed in the 1990s and used
in~\cite{Burrows:2018ggt,Gennari:2017yez} assume spin-saturation in nuclei and thus are most applicable to closed shell nuclei. Starting from a NCSM, even a closed
shell nucleus is not spin-saturated; e.g. a fully converged calculation for $^4$He requires N$_{\rm max} \ge12$,  where $N_{\max}$ is defined as the maximum number of oscillator quanta
above the valence shell for that nucleus. Though it is expected that closed shell nuclei are almost
spin-saturated, this will certainly not be the case for nuclei with partially filled shells. 
In order to take into account the spin of the struck nucleon a formulation to extract one-body
scalar and spin-densities from {\it ab initio} OBDMs needs to be developed. The scalar density is the one used
in the `traditional' folding interactions. From the spin-dependent 
densities, expectation values of the scalar product
of spin $\bm \sigma_i$ of the struck target nucleon with the momenta inherent in the nonlocal one-body density
must be
evaluated consistently with the operator structure of the NN Wolfenstein amplitudes. A first step
in this direction was made in~\cite{Orazbayev:2013dua} using a simple model density. Here we present  
scalar and spin-dependent OBDMs extracted from NCSM calculations.

\section{Theoretical Framework}

The NN scattering amplitude $\overline M$ can be parameterized  according to
Wolfenstein~\cite{wolfenstein-ashkin} in terms of six  
linearly independent spin-momentum operators multiplied by scalar functions of three linearly 
independent momentum vectors. The three vectors are the momentum transfer $\bf q$, the total
momentum of the system ${\bm{\mathcal{K}}}_{NN}$, and the normal to the scattering plane ${\bf n}_{NN}$,
\begin{eqnarray}
\label{2.1.1}
\lefteqn{
\overline{M}(\bm{q},\bm{\mathcal{K}}_{NN},\epsilon)= } & &  \cr 
& & A(\bm{q},\bm{\mathcal{K}}_{NN},\epsilon)\textbf{1}\otimes\textbf{1}
	+ iC(\bm{q},\bm{\mathcal{K}}_{NN},\epsilon)(\bm{\sigma}^{(0)}\otimes\textbf{1} +
\textbf{1}\otimes\bm{\sigma}^{(i)})\cdot \hat{n} \cr
	&+&
M(\bm{q},\bm{\mathcal{K}}_{NN},\epsilon)(\bm{\sigma}^{(0)}\cdot\hat{n})\otimes(\bm{\sigma}^{(i)}\cdot\hat{n}) \cr
	&+&
(G(\bm{q},\bm{\mathcal{K}}_{NN},\epsilon)-H(\bm{q},\bm{\mathcal{K}}_{NN},\epsilon))(\bm{\sigma}^{(0)}\cdot\hat{q})\otimes(\bm{\sigma}^{(i)}\cdot\hat{q}) \cr
	&+&
(G(\bm{q},\bm{\mathcal{K}}_{NN},\epsilon)+H(\bm{q},\bm{\mathcal{K}}_{NN},\epsilon))(\bm{\sigma}^{(0)}\cdot\hat{\mathcal{K}})\otimes(\bm{\sigma}^{(i)}\cdot\hat{\mathcal{K}}) \cr
	&+&
D(\bm{q},\bm{\mathcal{K}}_{NN},\epsilon)\left[(\bm{\sigma}^{(0)}\cdot\hat{q})\otimes(\bm{\sigma}^{(i)}\cdot\hat{\mathcal{K}})+(\bm{\sigma}^{(0)}\cdot\hat{\mathcal{K}})\otimes(\bm{\sigma}^{(i)}\cdot\hat{q})\right], 
\end{eqnarray}
where the scalar functions ({$A, C, M, G, H$, and $D$}) are the Wolfenstein amplitudes. The amplitude $D$ is zero
on-shell due to parity conservation.
The momenta are given as
\begin{eqnarray}
\label{2.1.2}
	\hat{q} = \frac{\left(\bm{k}' - \bm{k} \right)}{\left| \bm{k}'-\bm{k} \right|} \cr
	 \widehat{\mathcal{K}} = \frac{(\bm{k}' + \bm{k})}{\left| \bm{k}'+\bm{k} \right|} \cr
	\hat{n} = \frac{\bm{\mathcal{K}}\times \bm{q}}{|\bm{\mathcal{K}}\times \bm{q}|}~,
\end{eqnarray}
{with $\bm k$ and $\bm k'$, the initial and final momentum of the projectile nucleon.}
A spin-dependent space-fixed (sf) nonlocal one-body density between an initial A-body wave function
$|\Psi\rangle$ and a final A-body wave function $|\Psi'\rangle$, is written as:
\begin{equation}
\label{2.1.3}
\left(\rho_{sf}\right)^{K_s}_{q_s}(\bm{r},\bm{r'}) = \lla \Psi' \left| \sum_{i=1}^A \delta^3(\bm{r_i}- \bm
r) \delta^3(\bm{r'_i}- {\bm r'})  \left(\bm{\hat \tau}_{(i)}\right)_{q_s}^{K_s} \right| \Psi \rra~,
\end{equation}
where $ \bm{r_i}$ and $\bm{r'_i}$ are the initial and final space coordinate of the particle $i$, and $\bm r, \bm r'$ are parameters. Here  $\bm{\hat \tau_{(i)}}$ is the one body spin operator acting on particle $i$, a spherical tensor of rank $K_s = 0, 1$. When $K_s = 0$, the spin operator becomes the identity operator. 
\begin{eqnarray}
\label{tau}
K_s = 0~&:~~~~\left({\bm{\hat\tau_{(i)}}}\right)^{0}_{0}  =& 1 \cr
 \cr
K_s = 1~&:~~~~\left({\bm{\hat\tau_{(i)}}}\right)^{1}_{0}  =&  \bm{\sigma}_z \cr
&:~~\left({\bm{\hat\tau_{(i)}}}\right)^{1}_{-1} =& \frac{1}{\sqrt{2}} \left( \bm{\sigma}_x - i\bm{\sigma}_y \right) \cr
&:~~~~\left({\bm{\hat\tau_{(i)}}}\right)^{1}_{1}  =& -\frac{1}{\sqrt{2}} \left( \bm{\sigma}_x + i\bm{\sigma}_y \right)~.
\end{eqnarray}


In order to remove the center-of-mass (c.m.) contribution, the non-local one-body density
matrix is evaluated in momentum space, where we can employ the scheme given in detail in
Ref.~\cite{Burrows:2017wqn}. As function of the momentum variables ${\bf p}$ and ${\bf p}'$
the one-body density matrix 
$\left(\rho_{sf}\right)^{K_s}_{q_s}(\bm{p},\bm{p'})$ reads
\begin{eqnarray}
\label{2.2.21}
\left(\rho_{sf}\right)^{K_s}_{q_s}(\bm{p},\bm{p'}) &=& \sum_{nljn'l'j'} \sum_{K_l=|l-l'|}^{l+l'} \sum_{k_l=-K_l}^{K_l} \sum_{Kk} \lla K_l k_l K_s q_s | K k \rra \cr
&\times &(-1)^{J'-M'} \threej{J'}{K}{J}{-M'}{k}{M} \mathcal{Y}_{K k}^{*l l'}(\hat{p},\hat{p}') \cr
&\times & (-1)^{-l} \hat{j}\hat{j'}(K_s+1) \hat{s} \hat{K_s} \hat{K_l} \ninej{l'}{l}{K_l}{s}{s}{K_s}{j'}{j}{K} R_{n'l'}(p') R_{nl}(p) \cr
&\times & (-i)^{l+l'} \lla A \lambda' J' \left|\left| (a^{\dagger}_{n'l'j'} \tilde{a}_{nlj})^{(K)} \right|\right| A \lambda J \rra~.
\end{eqnarray}
The term $(a^{\dagger}_{n'l'j'} \tilde{a}_{nlj})^{(K)}$ represents the single particle
transition operator of rank K, with $\hat K=\sqrt{2K+1}$, and the corresponding reduced
matrix elements are characterized 
by the initial and final total angular momenta, J and J', while the remaining
quantum numbers are summarized by $\lambda$, and $\lambda'$.
These reduced matrix elements are provided by NCSM calculations.


To obtain translationally invariant densities we introduce as variables  the momentum
transfer $\bm{q}$ and the total momentum $\bm{\mathcal{K}}$, 
\begin{eqnarray}
\label{2.2.22}
	\bm{q} &=& \bm{p}'-\bm{p} \cr
	\bm{\mathcal{K}} &=& \half (\bm{p}' + \bm{p})~.
\end{eqnarray}
The spin-dependent one-body density (SOBD) is then derived in the same fashion as outlined
in~\cite{Burrows:2017wqn}, and we arrive at
\begin{eqnarray}
\label{SODM}
&&\left(\rho_{sf}\right)^{K_s}_{q_s}(\bm{q},\bm{\mathcal{K}}) = \sum_{nljn'l'j'} \sum_{K_l=|l-l'|}^{l+l'} \sum_{k_l=-K_l}^{K_l} \sum_{Kk} \lla K_l k_l K_s q_s | K k \rra \cr
&&\times (-1)^{J'-M'} \threej{J'}{K}{J}{-M'}{k}{M} (-1)^{-l} \hat{j}\hat{j'}(K_s+1) \hat{s} \hat{K_s} \hat{K_l} \cr
&&\times\ninej{l'}{l}{K_l}{s}{s}{K_s}{j'}{j}{K} (-i)^{l+l'} \cr
& &\times \sum_{n_q,n_{\mathcal{K}},l_q,l_{\mathcal{K}}} \lla n_{\mathcal{K}} l_{\mathcal{K}}, n_q l_q : K_l | n' l', n l : K_l \rra_{d=1} R_{n_{\mathcal{K}} l_{\mathcal{K}}}(\mathcal{K}) R_{n_q l_q}(q)  \cr
&&\times \mathcal{Y}_{K_l k_l}^{*l_{\mathcal{K}} l_q}(\hat{q},\hat{\mathcal{K}})\lla A \lambda' J' \left|\left| (a^{\dagger}_{n'l'j'} \tilde{a}_{nlj})^{(K)} \right|\right| A \lambda J \rra~.
\end{eqnarray}
In order to determine the center of mass (c.m.) contribution, we follow the same procedure as in \cite{Burrows:2017wqn}.
The center of mass wavefunction is assumed to be entirely in the 0s ground state of the nucleus. This results in the term $\zeta_{c.m.}$ being 0 for the c.m. contribution. Thus, using the relative coordinates from Eq.~(\ref{2.2.22}) and c.m. decomposition given by
 $ \left| \Psi \rra = \left| \Psi_{int} \rra \times \left| \Psi_{c.m.} 0s \rra $ we can separate the c.m. contribution and obtain the translational invariant part of the density. We obtained the center of mass contribution to be the same as in the scalar density.

\section{Expectation Values of Spin Dependent One-Body Density}

In order to evaluate the scattering amplitude of Eq.~({\ref{2.1.1}), expectation values of operators
involving the struck target nucleon, represented by the scalar product of ${\bm
\sigma}^{(i)}$ with one of
the momentum vectors, must be calculated.  If this operator is the unit operator, the result is a 
nonlocal scalar density, as e.g. used to calculate the effective NA interaction in
Ref.~\cite{Burrows:2018ggt}. The term proportional to the Wolfenstein amplitude $A$ leads to the central
part of the effective potential and the one proportional to $C$ to its spin-orbit part when scattering
from a spin-zero nucleus is considered. In general, the terms containing the scalar product of ${\bm
\sigma}^{(i)}$ with the momentum vectors of Eq.~(\ref{2.1.2}) need to be evaluated in the c.m. frame of the
nucleus. In the following, we will show explicit expressions for the expectation value of 
${\bm \sigma}^{(i)} \cdot {\hat n}_{t.i}$, which is the momentum space representation of the spin-operator.
The subscript $t.i.$ indicates that we use target c.m. momenta. 
We define
\begin{eqnarray}
\label{3.2}
	\hspace*{-1cm}\mathcal {S}_{n}(\bf{r},\bf{r'}) &=& \lla \Phi' \left| \sum_{i=1}^A \delta^3(\bf{r_i}-{\bf r}) \delta^3({\bf r_i'}-{\bf r}) \left[ {\bm \sigma}^{(i)} \cdot \hat{n} \right]^{0}_{0} \right| \Phi\rra \cr
	&=& \lla \Phi' \left| \sum_{i=1}^A \delta^3({\bf r_i}-{\bf r}) \delta^3({\bf r_i'}-{\bf r}) \left[ {\bm{{\tau}_{(i)}}}^{K_s=1} 
	\cdot {\bm{\hat{n}_{t.i.}}}^1 \right]^{0}_{0} \right| \Phi\rra.
\end{eqnarray}
Here spherical components of the spin tensor $\bf{\tau}_{(i)}$ are used and coupled with 
components of $\mathbf{{\hat n_{t.i.}}}$ to a tensor of rank 0. 
We rewrite $\mathcal{S}_n(\bf{r},\bf{r'})$ in terms of variables $\bf{q}$
and $\bf{\mathcal{K}}$ 
following the same procedure as in Eq.~(\ref{SODM}).  The final
expression, after expanding the projection, becomes
\begin{eqnarray}
\label{2.3}
\lefteqn{
\mathcal {S}_n(\bm{q},\bm{\mathcal{K}}) = } \cr
& & \sum_{q_s} (-1)^{-q_s} \lla 1 q_s 1 -q_s | 0 0 \rra \frac{4\pi}{3} \sum_{nljn'l'j'} \sum_{K_l=|l-l'|}^{l+l'} \sum_{k_l=-K_l}^{K_l} \sum_{Kk} \lla K_l k_l 1 q_s | K k \rra \cr
& &(-1)^{J'-M'} \threej{J'}{K}{J}{-M'}{k}{M} (-1)^{-l} \hat{j}\hat{j'}(2) \hat{s} \hat{1} \hat{K_l} \ninej{l'}{l}{K_l}{s}{s}{1}{j'}{j}{K} \cr
& & (-i)^{l+l'} \sum_{n_q,n_{\mathcal{K}},l_q,l_{\mathcal{K}}} \lla n_{\mathcal{K}} l_{\mathcal{K}}, n_q l_q : K_l | n' l', n l : K_l \rra_{d=1} R_{n_{\mathcal{K}} l_{\mathcal{K}}}(\mathcal{K}) R_{n_q l_q}(q) \cr
& & \sum_{LN} \lla K_l k_l 1 q_s | L N \rra \sum_{w w'} \frac{1}{(4\pi)^2} \hat{l_q}\hat{l_{\mathcal{K}}}\hat{1}\hat{1}\hat{K_l}\hat{1} 
 \lla l_q 0 1 0 | w' 0 \rra \lla l_{\mathcal{K}} 0 1 0 | w 0  \rra  \cr
& &   \ninej{l_q}{1}{w'}{l_{\mathcal{K}}}{1}{w}{K_l}{1}{L} \mathcal{Y}_{LN}^{w' w}(\hat{q},\hat{\mathcal{K}}) 
\lla A \lambda' J' \left|\left| (a^{\dagger}_{n'l'j'} \tilde{a}_{nlj})^{(K)} \right|\right| A \lambda J \rra~.
\end{eqnarray}


\begin{figure}[!ht]
\centering
\includegraphics[scale=0.6]{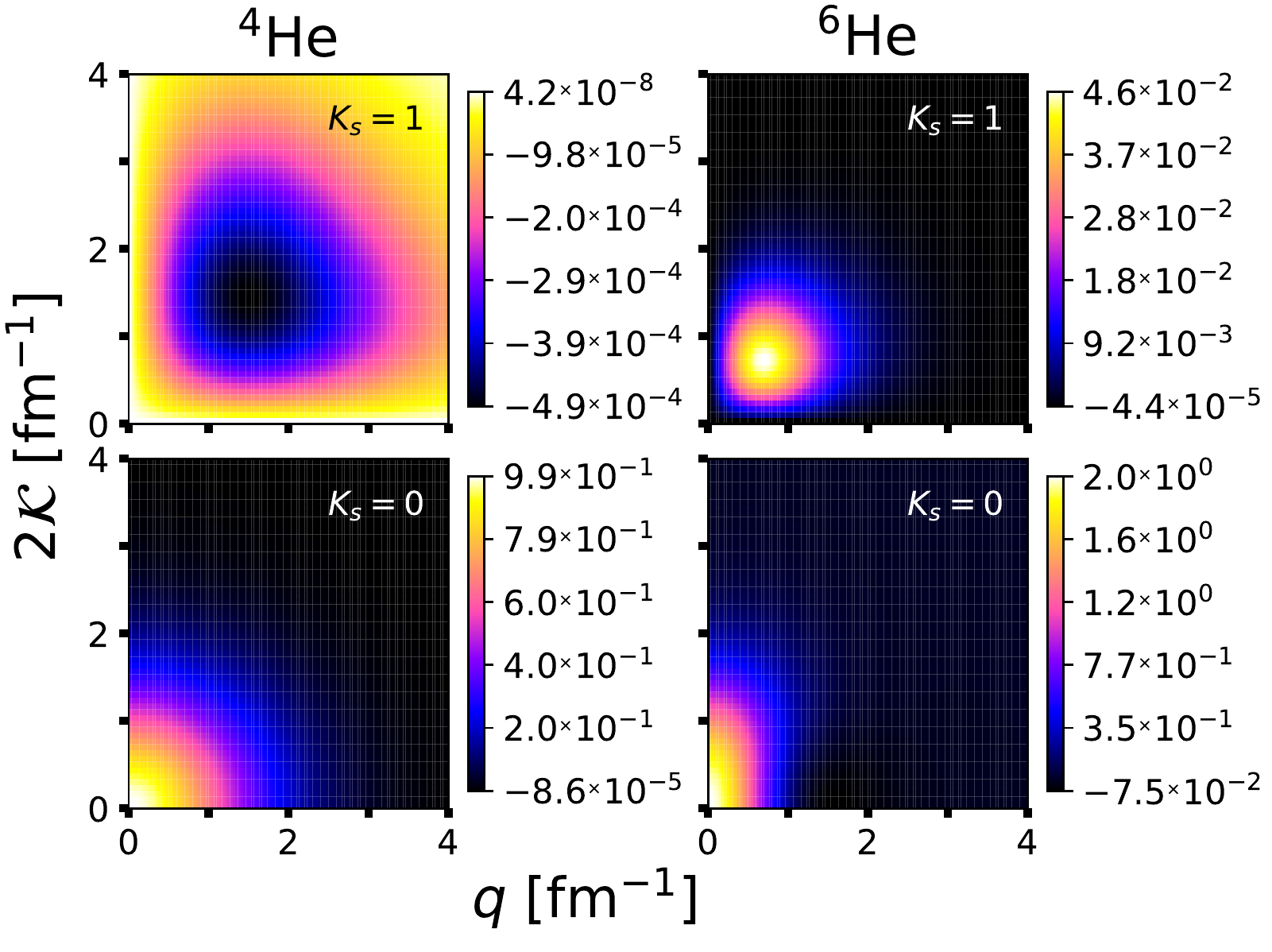}
  \caption[]{The expectation value of the scalar translationally
invariant nonlocal one-body density ($K_s=0$) and the spin-orbit
operator ($K_s=1$) as function of ${q}$ and ${\mathcal{K}}$, with the angle between them
fixed at $90^o$, and evaluated in the ground state.
All four graphs use one body reduced matrix elements from  NCSM calculations with $N_{\rm
max}$=18 and $\hbar \omega$=20~MeV based on the NNLO$_{\rm opt}$ interaction~\cite{Ekstrom13} for the neutron distribution of $^{4}$He (left panels) and $^{6}$He (right panels).}  
\label{fig1}
\end{figure}

While we concentrated in Eq.~(\ref{3.2}) on the expectation value of 
${\bm \sigma}^{(i)}\cdot {\hat n}$, expressions for the expectation values of the other scalar
products of ${\bm \sigma}^{(i)}$ with the unit vectors ${\hat q}$ and ${\hat {\mathcal K}}$
can be derived in a similar fashion. However, we notice that those expressions are
scalar products between a pseudo-vector and a vector, which are not invariant under parity
transformations. Thus, their expectation values between the ground states vanish, a fact
we numerically verified. 

For numerical studies, we only need to concentrate on the expectation values
represented by Eq.~(\ref{2.3}), which we want to call spin-orbit density.
We contrast the scalar nonlocal one-body density
($K_s$=0) with the density function given by the expectation value of
the spin-orbit operator ($K_s$=1) in the ground state of the nucleus, which we call
spin-orbit density.

\begin{figure}[!ht]
\centering
\includegraphics[scale=0.6]{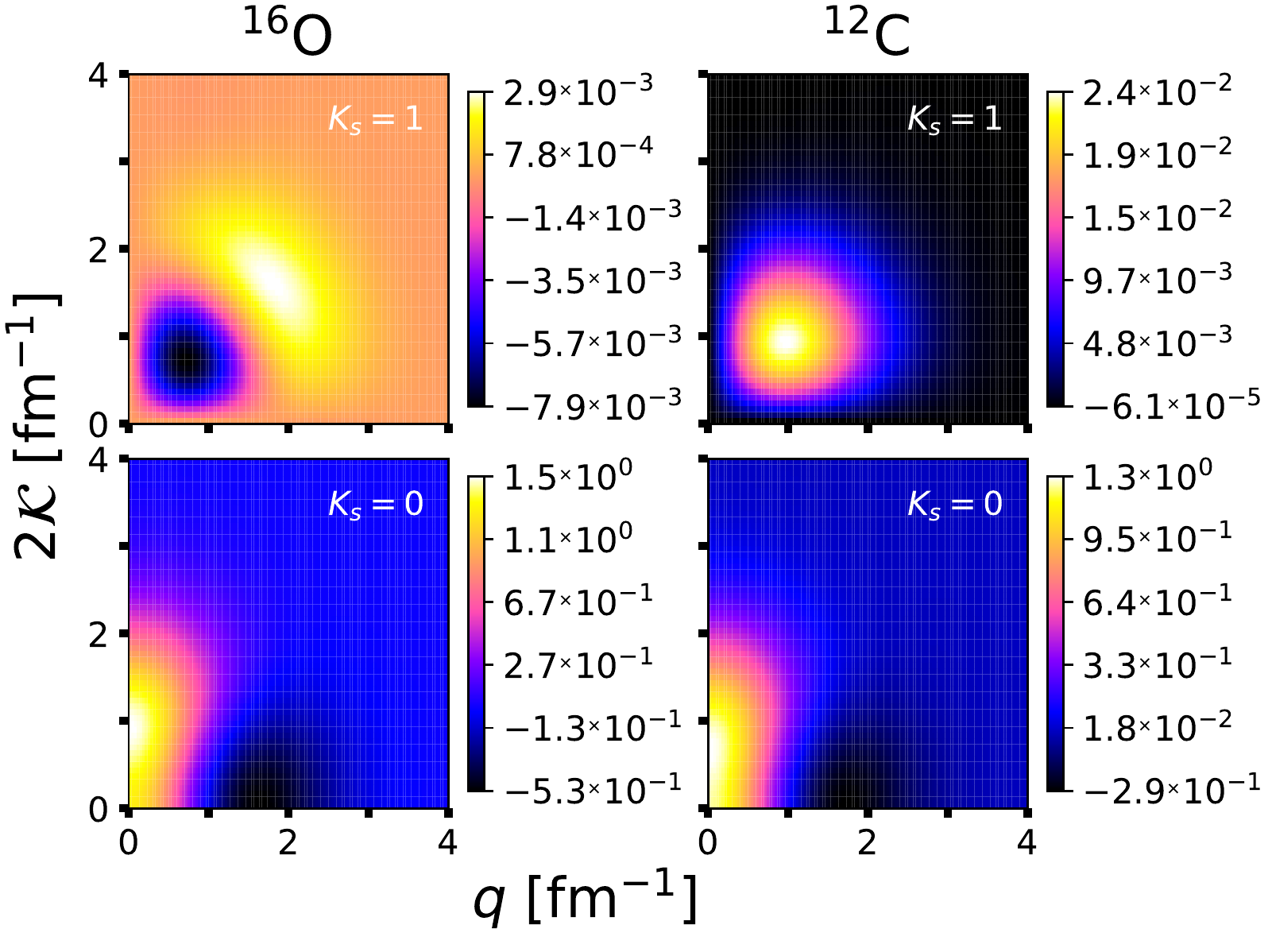}
  \caption[]{Same as Fig.~\ref{fig1} but comparing the neutron distribution of
$^{16}$O (left panels) and $^{12}$C (right panels).
The NCSM calculations were performed at $N_{\rm max}$=10 and $\hbar \omega$=20~MeV using
the NNLO$_{\rm opt}$ interaction~\cite{Ekstrom13}.}
\label{fig2}
\end{figure}

\begin{figure}[!ht]
\centering
\includegraphics[scale=0.6]{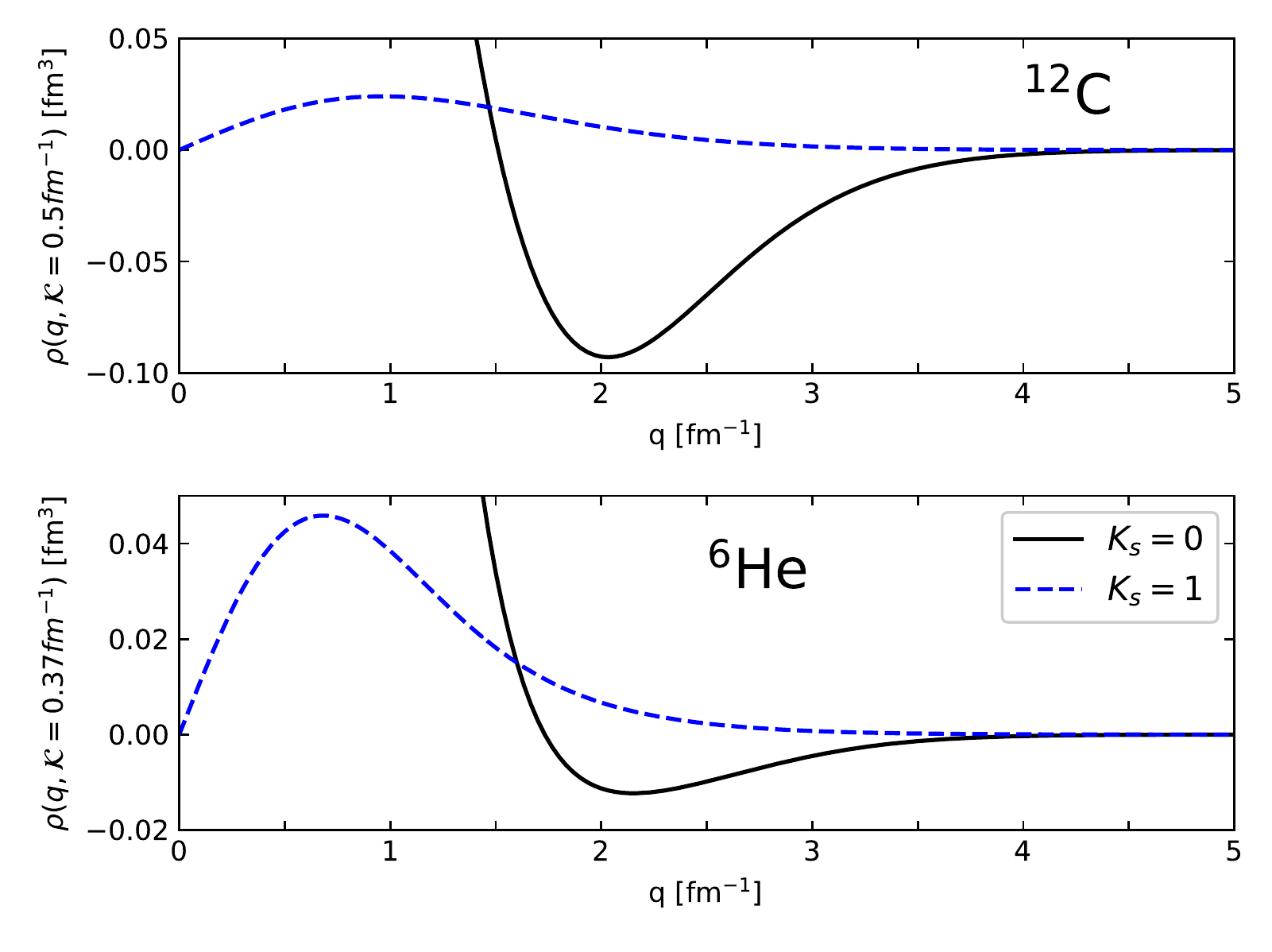}
  \caption[]{The expectation value of the scalar 
one-body density ($K_s=0$) (solid) and the spin-orbit density ($K_s=1$) (dashed) as function 
of the momentum transfer ${q}$ for the fix value of  ${\mathcal{K}}$=0.37~ fm$^{-1}$ 
(with the angle between them fixed at 90$^o$)
for $^6$He and ${\mathcal{K}}$=0.5~ fm$^{-1}$ for $^{12}$C. The NCSM calculations were performed at $N_{\rm max}$=10 and
$\hbar \omega$=20~MeV using the NNLO$_{\rm opt}$ interaction~\cite{Ekstrom13}.}
\label{fig3}
\end{figure}

The first example given in Fig.~\ref{fig1} shows the two cases for the closed-shell nucleus 
$^4$He and the open-shell nucleus $^6$He.
The expectation values of the scalar translationally
invariant nonlocal one-body density ($K_s=0$) and the spin-orbit
density ($K_s=1$) as function of ${\bf q}$ and 
${\bm{\mathcal{K}}}$, with the angle between them fixed at $90^o$, are evaluated in the ground state
of the corresponding nuclei.
All four graphs use as input one body reduced matrix elements from  NCSM calculations with $N_{\rm
max} = 18$ and $\hbar \omega = 20$ MeV based on the NNLO$_{\rm opt}$ interaction~\cite{Ekstrom13} 
 for the neutron distribution of $^{4}$He (left panels) and $^{6}$He (right panels).
For $^{4}$He, the spin-orbit contribution to the density is at least three orders of
magnitude smaller than the scalar part. It is still interesting to notice that its maximum
strength is at about $q = 1.5$ fm$^{-1}$, and $2\mathcal{K}=1.5$ fm$^{-1}$, away from  the
maximum value of the scalar part that is at $q=0$ and $\mathcal K=0$. For $^6$He, the
maximum value of the spin-orbit contribution is only two order of magnitude smaller than the
contribution from the scalar one. In this case, its maximum strength is at about $q = 0.7$
fm$^{-1}$, and $2\mathcal{K}=0.7$ fm$^{-1}$ away from $q=0$ and $\mathcal K=0$ where the
maximum of the scalar density is located. 
This does show the importance of the spin-orbit contribution to the density in an open-shell
nucleus like  $^6$He.

To further investigate the effect of the spin-orbit contribution to the density, we present the expectation value of the scalar translationally
invariant nonlocal one-body density ($K_s=0$) and the spin-orbit
operator ($K_s=1$) as function of ${\bf q}$ and ${\bf
\mathcal{K}}$, with the angle between them fixed at $90^o$ and evaluated in the ground state, for $^{16}$O and $^{12}$C, in Fig.~\ref{fig2}.
All four graphs use as input one body reduced matrix elements from  NCSM calculations with $N_{\rm max} = 18$ and $\hbar \omega = 20$ MeV based on the NNLO$_{\rm opt}$ interaction for the neutron distribution of $^{16}$O (left panels) and $^{12}$C (right panels).
These two sets of calculations are very interesting, since while the scalar density looks similar in strength and distribution over the momenta for both nuclei, the spin-orbit contribution looks quite different.  The maximum value of the spin-orbit contribution to the density is positive and about two order of magnitude smaller then the maximum value of the scalar density for $^{12}$C, at about $q = 1$ fm$^{-1}$ and $2\mathcal K = 1$ fm$^{-1}$,  while for $^{16}$O, the maximum value is negative and also about two order of magnitude smaller than the scalar one, at about $q = 0.7$ fm$^{-1}$ and $2\mathcal K = 0.7$ fm$^{-1}$, and has another positive  contribution distributed at different values of q and $\mathcal K$.

Looking at the absolute sizes of the scalar OBDs and the spin-orbit densities could lead to a
conclusion that the latter being at least two orders of magnitude smaller may render them negligible
in NA scattering calculations. For this consideration it is useful to recall the
on-shell condition ${\bf q}^2 + {\bm {\mathcal K}}^2 = 4 {\bf k_0}^2$ for NA scattering, where ${\bf
k_0}$ is the on-shell momentum in the NA frame and related to the c.m. scattering energy. The
on-shell conditions indicates that the maximum values on the scalar OBD close to the origin in
Figs.~\ref{fig1} and~\ref{fig2} are far off-shell in NA scattering. In Fig.~\ref{fig3} we compare
slices of the scalar and spin-orbit densities as function of $q$ for fixed, small ${\mathcal K}$
values for $^{12}$C and $^6$He. Here we see that for $q$-values of about 1.5~fm$^{-1}$ both
densities have the similar values, and the effect on scattering observables may very well be
visible. From Fig.~\ref{fig3}, we can also notice that the spin-density is relative constant in $^{12}$C, while for $^6$He has a maximum in momentum space at about 0.7 fm$^{-1}$, and goes to almost zero at about 2.5 fm$^{-1}$ making the strength focused in momentum space, that mean spread out in coordinate space.

Since the reduced matrix elements of the one body operators calculated in NCSM are dependent of the parameter $N_{\rm max}$, we performed a series of calculations with different values of $N_{\rm max}$. We present in Fig.~\ref{fig4} calculations of the expectation values of the  spin-orbit
operator ($K_s=1$) as function of ${\bm q}$ and 
${\bm{\mathcal{K}}}$, with the angle between them fixed at $90^o$ and evaluated in the ground state, for $^4$He and $^6$He, with $N_{\rm max} = 10$ and $N_{\rm max} = 18$.
All four graphs use as input one body reduced matrix elements from  NCSM calculations with  $\hbar \omega = 20$ MeV.
While the maximum contribution to the density is about $4.8 \times 10^{-4}$ for $^4$He with $N_{\rm max} = 10$, it increases only slightly to 
$4.9 \times 10^{-4}$ for  $N_{\rm max} = 18$.  Interestingly, for $^6$He, the strength of the spin-orbit contribution changes by about 15$\%$ (from 4.0$\times 10^{-2}$ to 4.6$\times 10^{-2}$) when increasing the NCSM space from $N_{\rm max} = 10$ to $N_{\rm max} = 18$. By increasing $N_{\rm max}$, the strength of the distribution also moves slightly towards lower values of $q$ and $\mathcal K$ for $^6$He. Since including more shells in the calculations drastically increasing the computation time, more calculations are underway in order to have a better understanding of how many shells are sufficient for converging results.

\begin{figure}[!ht]
\centering
\includegraphics[scale=0.6]{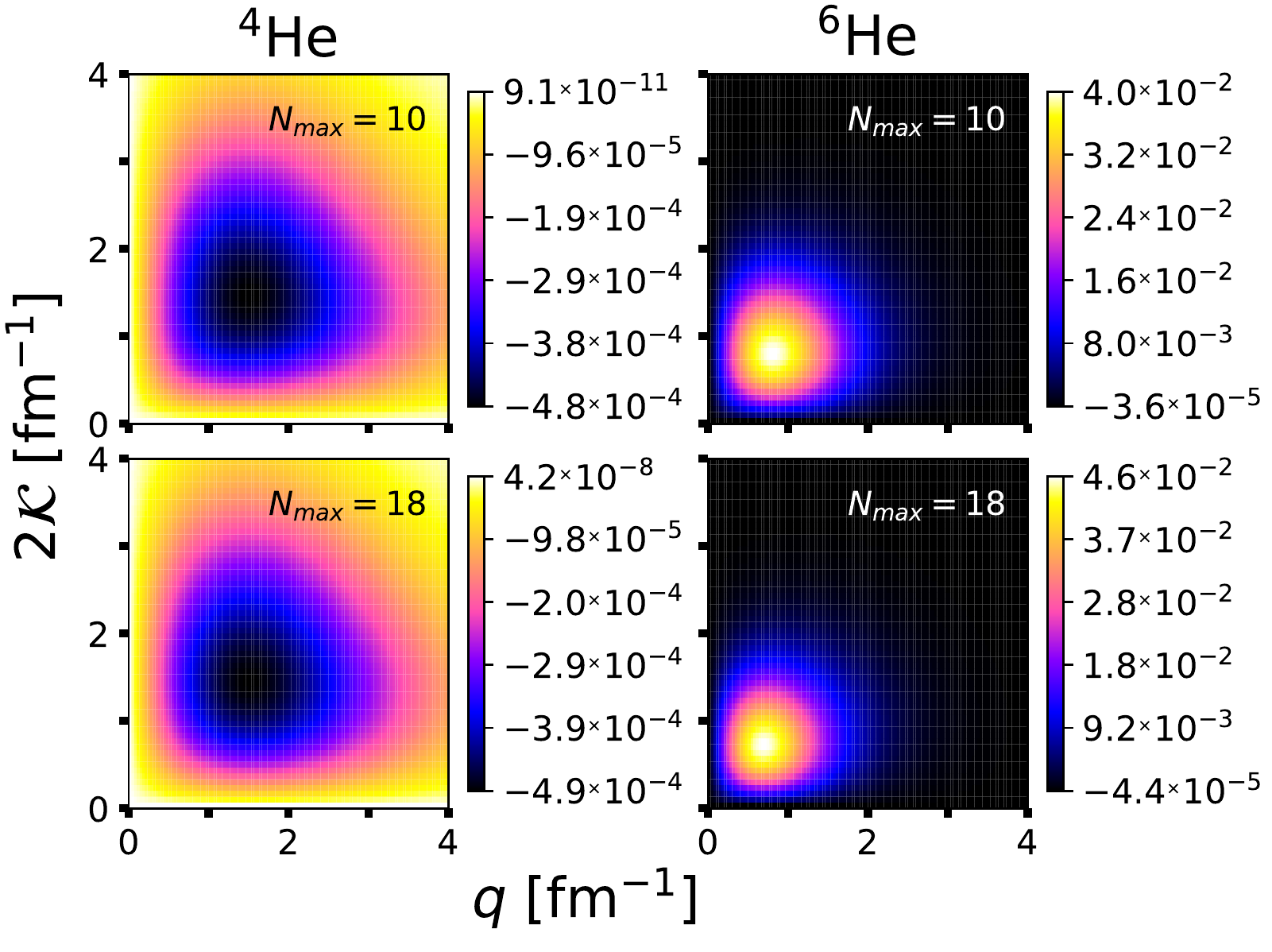}
  \caption[]{The neutron distribution of the spin-orbit operator expectation value, as a
function of the momenta $q$ and $\mathcal{K}$, with the angle between them fixed at
$90^o$, and evaluated in the ground state. The densities used as input the reduced matrix
elements from NCSM calculations with two different  values of $N_{\rm max} = 10$ and $18$.
All four calculations used input from NCSM with  $\hbar \omega = 20$ MeV and are based on
the NNLO$_{\rm opt}$ interaction~\cite{Ekstrom13} for $^4$He (left panels) and $^6$He (right panels).}
  \label{fig4}
\end{figure}


\section{Summary and Outlook}
We evaluated the expectation values of the scalar products of the spin of the struck target
nucleon with its three linearly independent target momenta, ${\hat q}$, ${\hat {\cal K}}$, and
${\hat n}$.  Only the expectation values of ${\bm \sigma}^{(i)} \cdot {\hat n}$, which
correspond to the momentum representation of the spin-orbit operator,
lead to a  non-vanishing contribution . Our calculations indicate that for open-shell nuclei like
$^6$He and $^{12}$C this spin-orbit density is considerably larger than for closed-shell nuclei
like $^4$He and $^{16}$O. Since the spin-orbit density for the stuck target nucleon enters
the NN scattering amplitude by means of the Wolfenstein amplitudes $C$ and $M$, one may
expect additional contributions to the traditional folding effective interaction for proton
scattering of spin-zero nuclei in the central potential through the Wolfenstein amplitude $C$ 
and in the spin-orbit potential through the  Wolfenstein amplitude $M$. Corresponding work
using the above developed expectation values is under way.

\section*{Acknowledgements}
Partial support for this work is given by the U.S. DoE under DE-FG02-93ER40756,
DE-SC0018223, DE-AC02-05CH11231, and the U.S. NSF under OIA-1738287, ACI-1713690,
OCI-0725070, and ACI-1238993. G.P acknowledges the support from the Ohio University Zanesville.

\bibliographystyle{h-physrev5}
\bibliography{clusterpot,denspot,ncsm}

\begin{thebibliography}{10}

\bibitem{Navratil:2000ww}
P.~Navratil, J.~P. Vary, and B.~R. Barrett,
\newblock Phys. Rev. Lett. {\bf 84}, 5728 (2000), arXiv:nucl-th/0004058.

\bibitem{Navratil:2000gs}
P.~Navratil, J.~P. Vary, and B.~R. Barrett,
\newblock Phys. Rev. {\bf C62}, 054311 (2000).

\bibitem{Roth:2007sv}
R.~Roth and P.~Navratil,
\newblock Phys. Rev. Lett. {\bf 99}, 092501 (2007), arXiv:0705.4069.

\bibitem{BarrettNV13}
B.~Barrett, P.~Navr\'{a}til, and J.~Vary,
\newblock Prog. Part. Nucl. Phys. {\bf 69}, 131 (2013).

\bibitem{Stumpf:2015lma}
C.~Stumpf, J.~Braun, and R.~Roth,
\newblock Phys. Rev. {\bf C93}, 021301 (2016), arXiv:1509.06239.

\bibitem{Weppner:2000fi}
S.~P. Weppner, O.~Garcia, and C.~Elster,
\newblock Phys. Rev. {\bf C61}, 044601 (2000).

\bibitem{Koning:2003zz}
A.~Koning and J.~Delaroche,
\newblock Nucl.Phys. {\bf A713}, 231 (2003).

\bibitem{Furumoto:2019anr}
T.~Furumoto, K.~Tsubakihara, S.~Ebata, and W.~Horiuchi,
\newblock Phys. Rev. {\bf C99}, 034605 (2019).

\bibitem{Chinn:1993zz}
C.~R. Chinn, C.~Elster, and R.~M. Thaler,
\newblock Phys.Rev. {\bf C48}, 2956 (1993).

\bibitem{Elster:1989en}
C.~Elster, T.~Cheon, E.~F. Redish, and P.~C. Tandy,
\newblock Phys. Rev. {\bf C41}, 814 (1990).

\bibitem{Crespo:1990zzb}
R.~Crespo, R.~C. Johnson, and J.~A. Tostevin,
\newblock Phys. Rev. {\bf C41}, 2257 (1990).

\bibitem{Arellano:1990xu}
H.~F. Arellano, F.~A. Brieva, and W.~G. Love,
\newblock Phys. Rev. {\bf C41}, 2188 (1990),
\newblock [Erratum: Phys. Rev.C42,1782(1990)].

\bibitem{Burrows:2018ggt}
M.~Burrows {\em et~al.},
\newblock Phys. Rev. {\bf C99}, 044603 (2019), arXiv:1810.06442.

\bibitem{Gennari:2017yez}
M.~Gennari, M.~Vorabbi, A.~Calci, and P.~Navratil,
\newblock Phys. Rev. {\bf C97}, 034619 (2018), arXiv:1712.02879.

\bibitem{Orazbayev:2013dua}
A.~Orazbayev, C.~Elster, and S.~P. Weppner,
\newblock Phys. Rev. {\bf C88}, 034610 (2013), arXiv:1305.6964.

\bibitem{wolfenstein-ashkin}
L.~Wolfenstein and J.~Ashkin,
\newblock Phys. Rev. {\bf 85}, 947 (1952).

\bibitem{Burrows:2017wqn}
M.~Burrows {\em et~al.},
\newblock Phys. Rev. {\bf C97}, 024325 (2018), arXiv:1711.07080.

\bibitem{Ekstrom13}
A.~Ekstr{\"o}m {\em et~al.},
\newblock Phys. Rev. Lett. {\bf 110}, 192502 (2013).

\end{thebibliography}



\end{document}